\documentclass[12pt]{article}            

\usepackage{amsmath,amssymb,amsthm}
\usepackage{mathbbol,bbm}
\usepackage[twoside=false,dvips=true]{geometry}
\allowdisplaybreaks[4]

\newcommand{\Cx}{{\mathbb C}}

\newcommand{\Rl}{{\mathbb R}}

\newcommand{\idty}{\b 1}
\DeclareMathOperator{\id}{id}

\newcommand{\co}[1]{\textsf{#1}}

\DeclareMathOperator*{\tr}{Tr}
\newcommand{\<}{\langle}
\renewcommand{\>}{\rangle}
\providecommand{\abs}[1]{\lvert#1\rvert}
\providecommand{\norm}[1]{\lVert#1\rVert}

\renewcommand{\b}[1]{\mathbf{#1}}
\newcommand{\bi}[1]{\boldsymbol{#1}}
\renewcommand{\c}[1]{\mathcal{#1}}

\newcommand{\s}[1]{\mathsf{#1}}
\renewcommand{\r}[1]{\mathrm{#1}}

\newcommand{\ket}[1]{|\,#1\rangle}
\newcommand{\bra}[1]{\langle#1\,|}

\newcommand{\uu}{{x_{\uparrow\uparrow}}}
\newcommand{\ud}{{x_{\uparrow\downarrow}}}
\newcommand{\du}{{x_{\downarrow\uparrow}}}
\newcommand{\dd}{{x_{\downarrow\downarrow}}}

\newtheorem{theo}{Theorem}
\newcommand{\R}{\r{Re}}
\newcommand{\I}{\r{Im}}
\DeclareMathOperator*{\loplus}{\mbox{\Large\mbox{$\oplus$}}}

\begin{document}

\begin{center}
\textbf{{\Large Finite size mean-field models}\footnote{This paper is dedicated to
A.~Verbeure on the occasion of his 65th anniversary.}  \\[6pt]
{\large M.~Fannes}\footnote{E-mail: mark.fannes@fys.kuleuven.be} 
and
{\large C.~Vandenplas}\footnote{E-mail: caroline.vandenplas@fys.kuleuven.be} \\[6pt]
Instituut voor Theoretische Fysica, K.U.Leuven \\ 
Celestijnenlaan~200D, B-3001 Heverlee, Belgium}
\end{center}
\bigskip

\noindent
\textbf{Abstract:}
We characterize the two-site marginals of exchangeable states of a system of
quantum spins in terms of a simple positivity condition. This result is used
in two applications. We first show that the distance between two-site marginals of
permutation invariant states on $N$ spins and exchangeable states is of
order $1/N$. The second application relates the mean ground state energy of a
mean-field model of composite spins interacting through a product pair interaction 
with the mean ground state energies of the components.    
\medskip

\section{Introduction}

The mean-field approximation is a very common approach in statistical mechanics. It 
consists in replacing suitably chosen parts of the interaction by their expectation
values. This generally simplifies the problem but leads to non-linear self-consistent
equations for the dynamics and the equilibrium states. This kind of approximation 
often leads to reasonable results in regimes where the interactions are rather weak. 
In other cases, the self-consistency equations may induce artificial phase 
transitions~\cite{Rei}. 

A characteristic feature of the most basic version of the approximation is that every
particle interacts in the same way with every other particle. Therefore the
Hamiltonian and the ground and equilibrium states have a huge symmetry: particles can
be arbitrarily permuted. By mean-field models we here mean quantum spin systems 
which exhibit this kind of symmetry. There is a vast literature on the subject dealing
both with the structure of states and dynamical maps \cite{Fuc,Hud,Sto}.
We briefly recall some essential notions and results.  
    
A state $\omega$ of a system of $N$ identical spin-$(d\!-\!1)/2$ particles is 
determined by a density matrix $\rho \in \c M_{d^N}(\Cx) = \otimes_N \c M_d(\Cx)$, 
where $\c M_d(\Cx)$ denotes the complex matrices of dimension $d$
\begin{equation*}
 \omega(A) = \tr \rho\, A,\quad \text{for } A \in \otimes_N \c M_d(\Cx).
\end{equation*}
An $N$-particle state is \co{symmetric} if it is invariant under permutations of 
the particles, i.e., if 
\begin{equation}
 U_\pi(\ket{\Psi_1} \otimes \cdots \otimes \ket{\Psi_N}) = 
 \ket{\Psi_{\pi(1)}} \otimes \cdots \otimes \ket{\Psi_{\pi(N)}}\quad
 \text{where } \ket{\Psi_i} \in \Cx^d, 
\label{rep}
\end{equation} 
then
\begin{equation*}
 \omega(A) = \omega(U_\pi AU^*_\pi) \quad \text{for every permutation } 
 \pi \text{ of } \{1,\ldots,N\} \text{ and } A \in \otimes_N \c M_d(\Cx).
\end{equation*}
In terms of the density matrix $\rho$ of $\omega$,
\begin{equation*}
 \rho = U^*_\pi\rho\, U_\pi.
\end{equation*}

Consider a system of $(N+M)$ particles with a symmetric state $\omega$. 
For each subsystem of $N$ particles the marginals of $\omega$ are symmetric 
$N$-particle states $\omega_N$  
\begin{equation*}
 \omega_N(A) := \omega\bigl( A \otimes (\otimes_M\, \idty) \bigr)\quad
 \text{for every } A\in \otimes_N \c M_d(\Cx). 
\end{equation*}
Note that, due to the symmetry of $\omega$, only the number of spins in a 
subsystem matters and not the precise sites on which the subsystem lives. 
The density matrices $\rho_N$ associated with these states are obtained 
by taking partial traces of the density matrix $\rho$ that defines $\omega$ 
\begin{equation*}
 \rho_N = \r{Tr}_M\, \rho = 
 \sum_{(i_1,\ldots, i_M)} \bigl( \id \otimes \ket{e_{i_1} \cdots e_{i_M}} 
 \bra{e_{i_1} \cdots e_{i_M}} \bigr)\, (\rho), 
\end{equation*}
where $\{e_i\}_{i=0}^{d-1}$ is a basis of $\Cx^d$. The converse is not true, 
a symmetric $N$-particle state $\omega$ cannot always be extended
to a symmetric $(N+M)$-particle state. For example, consider the pure two-qubit state determined by $\ket{\Psi} = \frac{1}{\sqrt{2}} (\ket{01}+\ket{10})$. 
This state  is symmetric but has no symmetric extension to three qubits
\cite{Wer}.
  
If we want a symmetric state to have a symmetric extension to an arbitrarily large
system, we have to impose the stronger condition of exchangeability.
A state $\omega$ on $\otimes_N \c M_d(\Cx)$ is called \co{exchangeable} 
if it admits for any $M>0$ a symmetric extension $\omega_{(N+M)}$ to 
$\otimes_{N+M} \c M_d(\Cx)$. Exchangeability is a quite strong condition, as we 
see in the following quantum version of de~Finetti's theorem \cite{Fin,Hew}.

\begin{theo}
\label{De Finetti}
 If $\omega$ is an exchangeable state on $\otimes_N \c M_d(\Cx)$, then
 \begin{equation*}
  \omega = \int_{\c S_d} \!\r d\mu(\sigma)\, \otimes_N \sigma
 \end{equation*}
 where $\c S_d$ denotes the state space of $\c M_d(\Cx)$ and $\mu$ is a probability 
 measure on $\c S_d$.
\end{theo}

The exchangeable states are mixtures of symmetric product states which 
implies that they are non-entangled and so only classical correlations are
possible~\cite{Nie}. The inverse implication is not true, not every 
symmetric separable state is exchangeable. 
Consider, for instance, two density matrices $\rho,\sigma\in\c M_d(\Cx)$, then the state 
associated with $\frac{1}{2} (\rho\otimes\sigma + \sigma\otimes\rho)$ is symmetric and 
separable on $\c M_d(\Cx)\otimes\c M_d(\Cx)$ but generally not exchangeable. 

\section{Two-site marginals of exchangeable states}

We want to characterize the exchangeable states on two particle systems with 
$d$ degrees of freedom.
 
\begin{theo}
\label{characterisation B-E}
 A symmetric state $\omega$ on $\c M_d(\Cx) \otimes \c M_d(\Cx)$ is exchangeable 
 iff 
 \begin{equation*}
  \omega(B\otimes B) \ge 0\quad \text{for all } B \in \c M_d^{\r h}(\Cx) 
 \end{equation*}
 where $\c M_d^{\r h}(\Cx)$ denotes the complex Hermitian matrices of dimension~$d$.
\end{theo}

\begin{proof}
If $\omega$ is an exchangeable state two-particle state, then by 
theorem~\ref{De Finetti} we have that
\begin{equation*}
 \omega(B\otimes B) = \int_{\c S_d} \!\r d\mu(\sigma)\, \sigma(B)^2 \ge 0\quad 
 \text{for every } B \in \c M_d^{\r h}(\Cx).
\end{equation*} 
The remaining of the proof is postponed until section \ref{The quantum case}. 
\end{proof}

In order to prove the inverse direction we use the polar 
cone theorem to invert the role of states and observables.
So, instead of proving that $\omega$ is exchangeable if $\omega(B\otimes B) 
\ge 0$ for every 
$B \in \c M_d^{\r h}(\Cx)$, we will prove that a flip-invariant, hermitian 
$A \in \c M_d(\Cx) \otimes \c M_d(\Cx)$ is a positive combination 
of $B_\alpha \otimes B_\alpha$, $B_\alpha \in \c M_d^{\r h}(\Cx)$, 
if $\tr (\sigma\otimes\sigma\,A) \ge 0$ for every density matrix $\sigma \in \c S_d$.

More explicitly, given a real Hilbert space $\c H$ and a set $C \subset \c H$, the cone 
\begin{equation*} 
 C^* := \{ y \mid \<x,y\> \ge 0\ \text{for every } x \in C\},
\end{equation*} 
is called the polar cone of $C$.

\begin{theo}
Let $\c H$ be a real Hilbert space and $C$ a subset of $\c H$, then 
\begin{equation*}
 (C^*)^* = \overline{\r{Cone}(C)},
\end{equation*} 
where $\overline{\r{Cone}(C)}$ denotes the closure of the cone generated by $C$. 
\end{theo}

Let $F$ be the flip operator on $\Cx^d \otimes \Cx^d$
\begin{equation*}
 F(\varphi \otimes \psi) := \psi \otimes \varphi. 
\end{equation*}
We consider the real subspace $\c K$ of the complex hermitian matrices of dimension $d^2$
which commute with $F$ and equip $\c K$ with the trace scalar product 
\begin{equation*}
 \<\cdot,\cdot\>: \c M_{d^2}^{\r h}(\Cx) \times \c M_{d^2}^{\r h}(\Cx) \to \Cx:
 (A_1,A_2) \mapsto \tr A_1\,A_2.
\end{equation*}
We choose $C$ to be the set of all symmetric two-site product states determined by 
density matrices on $\Cx^d$ 
\begin{equation*}
 C = \bigl\{ \rho\otimes\rho \,\bigm|\, \rho\in\c M_d(\Cx),\ \text{ $\rho$
 is a density matrix} \bigr\}.
\end{equation*}
It is then enough to prove that the polar cone of $C$ is the closed cone $C^*$ generated 
by  
\begin{equation*}
 \bigl\{ B\otimes B \,\bigm|\, B\in\c M_d^{\r h}(\Cx) \bigr\}\ \cup\  
 \bigl\{ L \,\bigm|\, L\in\bigl( \c M_d(\Cx) \otimes\c M_d(\Cx) \bigr)^{\r h},\ 
 L \ge 0 \text{ and } L\,F = F\,L \bigr\}
\end{equation*} 
where $L\ge 0$ means that $L$ is a positive semi-definite matrix. Indeed, applying 
the polar cone theorem, we get 
\begin{align*}
 C^{**} 
 &= \bigl\{ \rho \,\bigm|\, \tr \rho(B\otimes B) \ge 0, B \in \c M^{\r h}_d(\Cx) 
 \text{ and } \tr\rho L\ge 0, L\ge 0\} \\
 &= \overline{\r{Cone}(C)} 
 = \r{Cone} \bigl( \{\rho\otimes\rho \mid \rho\in \c S_d\} \bigr).
\end{align*}

We shall first prove the analogous result for the classical case, that is 
when we replace the matrix algebra $\c M_d(\Cx)$ by the diagonal matrices of 
dimension $d$ and states by probability measures on the relevant configuration 
space. 

\subsection{A classical intermezzo}
\label{classical case}

Let $\Omega$ be a finite set and 
\begin{equation*}
 C := \{\mu\times\mu \mid\, \mu\text{ is a probability measure on $\Omega$} \}
\end{equation*} 
then
\begin{align*}
 C^* = \{f: \Omega\times\Omega\to\Rl \mid 
 &f(x,y)=f(y,x) \text{ and } \\
 &(\mu\times\mu)(f) \ge 0 \text{ for all measures } \mu \text{ on } \Omega\}.
\end{align*}

The aim is to show that the cone $C^*$ is generated by functions of the form
$f_1 + f_2$ where
\begin{itemize}
\item[i)]
\vspace{-\parskip}
 $f_1 \ge 0$ and $f_1(x,y) = f_1(y,x)$
\item[ii)]  
 $f_2 = g\times g$ with $g:\Omega\to\Rl$
\end{itemize}

Fix $f$ in the interior of $C^*$. By subtracting from $f$ a suitably chosen 
non-negative symmetric function, we can arrange to have a strictly positive 
measure $\mu_0$ on $\Omega$ 
such that 
\begin{equation}
 (\mu_0\times\mu_0)(f) = 0
 \qquad\text{and}\qquad
 (\mu\times\mu)(f) \ge 0 \quad\text{for all measures } \mu.
\label{posmeas}
\end{equation}

Let $\mu_0$ now be a measure as in~(\ref{posmeas}). For any $\tau$, 
a sufficiently small real functional on $\Omega$, 
$\mu_0+\tau$ is non-negative on $\Omega$. Therefore, by assumption,
\begin{equation}
 \bigl( (\mu_0+\tau) \times (\mu_0+\tau) \bigr) (f) \ge 0.
\label{2}
\end{equation}
As $(\mu_0\times\mu_0)(f) = 0$, this can only hold if
\begin{center}
\vspace{-\parskip}
 $(\mu_0\times\tau)(f) = 0$ for all choices of $\tau$ on $\Omega$
\end{center}
In this case,  
condition~(\ref{2}) translates into
\begin{equation}
 (\tau\times\tau)(f) \ge 0,\quad \text{for all } \tau.
\label{5}
\end{equation}
As the matrix $F := [f(x,y)]$ is real and equal to its transpose, (\ref{5})
amounts to requiring that $F$ be semi-definite positive. But then there exist $c_j(x)$
such that
\begin{equation*}
f(x,y)=[F]_{x,y} = \sum_j [c_j(x)\, c_j(y)],
\end{equation*} 
proving our statement.

\subsection{The quantum case\label{The quantum case}}
\label{quantum case}

\begin{proof}[Proof of second part of Theorem~\ref{characterisation B-E}]

We have now $C:=\{\rho\otimes\rho \mid \rho \text{ is a density matrix in }\\\c M_d(\Cx)\}$ and 
\begin{equation*}
 C^* := \{ A\in\c M^{\r h}_{d^2}(\Cx) \mid A\,F = F\,A \text{ and } 
 \tr A(\rho\otimes \rho) \ge 0 \text{ $\forall$ density matrices } 
 \rho\in\c M_d(\Cx)\}.
\end{equation*}
The aim is to prove that the cone $C^*$ is generated by matrices of the form
$A_1 + A_2$ with
\begin{itemize}
\item[i)]
\vspace{-\parskip}
 $A_1 \ge 0$ and $A_1\,F = F\,A_1$.
\item[ii)]  
 $A_2 = B\otimes B$ with $B\in\c M^{\r h}_d(\Cx)$.
\end{itemize}

As in the previous section we fix $A$ in the interior of 
$C^*$ and subtract from $A$ a positive semi-definite 
matrix to have an invertible density matrix $\rho_0$ such that
\begin{equation*}
 \tr A(\rho_0\otimes\rho_0) = 0
 \qquad\text{and}\qquad
 \tr A(\rho\otimes\rho) \ge 0 \text{ for all density matrices } \rho.
\end{equation*}
For any choice of $B \in \c M^{\r h}_d(\Cx)$, with $\norm B$ sufficiently 
small, $\rho_0+B$ is still positive semi-definite and so
\begin{equation}
 \tr A \bigl( (\rho_0+B) \otimes (\rho_0+B) \bigr) \ge 0.
\label{condition}
\end{equation}
As $\tr A(\rho_0\otimes\rho_0) = 0$, this can only hold if
\begin{equation*}
 \tr (\rho_0\otimes B)A = 0\text{ for every } B\in\c M^{\r h}_d(\Cx).
\end{equation*}
In this case, condition~(\ref{condition}) translates into
\begin{equation}
 \tr (B\otimes B)A \ge 0 \text{ for every } B\in\c M^{\r h}_d(\Cx).
\label{cond}
\end{equation}

We now extend the argument for the classical case, see section~\ref{classical case} 
to the quantum case. Therefore we introduce real linear maps, 
\begin{align*}
 &V_d: \c M^{\r h}_d(\Cx) \rightarrow \c H\quad  \text{and} \\ 
 &M_d: \bigl(\c M_d(\Cx) \otimes \c M_d(\Cx) \bigr)^{\r h} \rightarrow 
 \c B(\c H),
\end{align*} 
where $\c H$ is a suitably chosen real Hilbert space and $\c B(\c H)$ denotes the 
linear operators on that space such that:
\begin{itemize}
\item[i)]
 $V_d$ and $M_d$ are one-to-one and onto.
\item[ii)] 
 For every $A\in C^*$, $M_d(A)$ is positive, this will follow from condition(\ref{cond}), 
 $M_d(A)^{\s T} = M_d(A)$ and $\tr A(B\otimes B) = \bra{V_d(B)} M_d(A) \ket{V_d(B)}$.
\item[iii)]  
 $M^{-1}_d \bigl( \ket{\tau}\bra{\tau} \bigr) = V_d^{-1}(\tau) \otimes V_d^{-1}(\tau)$. 
\end{itemize}

With these maps we can prove that $A = \sum_\alpha B_\alpha \otimes B_\alpha$. Indeed,
as
\begin{equation*}
 \tr A(B\otimes B) = \bra{V_d(B)} M_d(A) \ket{V_d(B)} \ge 0 \quad\text{for every }
 B\in\c M^{\r h}_d(\Cx)
\end{equation*}
and $V_d$ is onto, we get that $M_d(A)$ is positive or $M_d(A) = \sum_\alpha 
\ket{\tau_\alpha} \bra{\tau_\alpha}$. Now, because $M_d$ is one-to-one and using 
property~(iii) above, we have 
\begin{equation*}
 A = M_d^{-1} \Bigl(\sum_\alpha \ket{\tau_\alpha}\bra{\tau_\alpha} \Bigr) 
 = \sum_\alpha M_d^{-1} \bigl( \ket{\tau_\alpha} \bra{\tau_\alpha} \bigr) 
 = \sum_\alpha V_d^{-1}(\tau_\alpha) \otimes V_d^{-1}(\tau_\alpha),
\end{equation*} 
proving our statement. Constructing the maps $V_d$ and $M_d$ and verifying 
their properties is rather tedious. We therefore provide the details 
separately in appendices A--C.
\end{proof}

\section{Finite size symmetric states}

In this section we focus on the distance between the two-site marginal
of an $N$-particle symmetric state and the two-site exchangeable states.
Let $\c S^N$ be the set of symmetric states $\omega$ of two particles 
which have a symmetric extensions to $N$ sites and let $\c S^\infty$ be the 
exchangeable two-particle states. Obviously,
\begin{equation*}
 \c S^2 \supset \c S^3 \cdots \supset \c S^N \supset\c S^{N+1} \cdots \supset
 \c S^\infty.
\end{equation*}
The sets $\c S^N$ are closed and convex in the state space of 
$\c M_d(\Cx) \otimes \c M_d(\Cx)$ for all $N=2,3,\ldots$. We can now 
wonder about the distance of $\c S^N$ to the exchangeable states $\c S^\infty$
\begin{align}
 \r d(\c S^N,\c S^\infty)
 &=\max_{\omega \in \c S^N} \r d(\omega, \c S^\infty) 
 = \max_{\omega \in \c S^N}\ \min_{\omega' \in \c S^\infty} \norm{\omega 
 - \omega'} 
\nonumber \\
 &= \max_{\omega \in \c S^N}\ \min_{\omega' \in \c S^\infty} \tr \abs{\rho 
 - \rho'},
\label{dis}
\end{align}
where $\rho$ and $\rho'$ are the density matrices corresponding to the two-site
states $\omega$ and $\omega'$. We know that for $N \rightarrow \infty$, 
this distance vanishes, but we are interested in the behaviour with $N$.
An upper bound of the order $1/\sqrt N$ was obtained in~\cite{Koe}. 
Such bounds yield a measure of the maximal entanglement of states in $\c S^N$. 
For a detailed analysis of a model, see e.g.~\cite{Duk}.

A possible approach to this question is to use the information on the
structure of symmetric states that can be obtained from group theory.
The decomposition of the natural representation of the permutation group $\c
S_N$ of
a set of $N$ elements on $\bigl(\Cx^d\bigr)^{\otimes N}$ given in~(\ref{rep})
in irreducible representations is a highly non-trivial achievement of group
theory~\cite{Ham}. 
The result is that the irreducible representation of $\c S_N$ are
labeled by \co{standard Young tableaux} $T$. The irreducible
representation corresponding to $T$ has dimension $d(T)$ and occurs with a 
multiplicity $m(T)$, both $d$ and $m$ are explicitly known, moreover, $d$
depends on $N$ and $m$ on $N$ and $d$. Hence, 
there is a decomposition
\begin{equation}
 \bigl(\Cx^d\bigr)^{\otimes N}  
 = \loplus_T\,  \Cx^{m(T)} \otimes \Cx^{d(T)}. 
\label{young}
\end{equation}
Any symmetric $N$-particle density matrix is then of the form
\begin{equation}
 \rho = \loplus_T\, c(T)\, \rho_T \otimes \idty
\label{rho}
\end{equation}
where $\rho_T$ is a density matrix on $\Cx^{m(T)}$ and $c(T)$ are suitably
chosen non-negative normalization coefficients. In order to estimate the
distance~(\ref{dis}) we can compute the two-site marginals of a state
determined by a pure $\rho_T$ in~(\ref{rho}) and estimate its distance from
the exchangeable states. Such a computation is, however, rather involved. We
nevertheless sketch an example of the computation for the case $d=2$. 

Considering $\Cx^2$ as the state space of a single spin-1/2 particle, the
decomposition~(\ref{young}) is nothing else than the standard decomposition
of a system of $N$ spin-1/2 particles according to total spin. Any value of
the spin in $\{0,1,\ldots, N/2\}$ for even $N$ and $\{1/2, 3/2,\ldots, N/2\}$
for odd $N$ occurs. Let us simplify the problem even further by choosing a
completely symmetric normalized vector $\Psi$ in $\bigl(\Cx^2\bigr)^{\otimes
N}$. We fix canonical basis vectors $|\uparrow\>$ and $|\downarrow\>$ in
$\Cx^2$, e.g.\ to the eigenstates of the $z$-component of the spin. A
natural basis of the completely symmetric subspace of 
$\bigl(\Cx^2\bigr)^{\otimes N}$ is then $\bigl\{ |n\> \,\bigm|\,
n=0,1,\ldots N \bigr\}$ where $|n\>$ is the normalized state obtained by
symmetrizing an elementary tensor with $n$ factors $|\uparrow\>$ and $N-n$
factors $|\downarrow\>$. Our vector $\Psi$ can then be written as
\begin{equation}
 \Psi = \sum_{n=0}^N \alpha_n\, |n\>,
\label{Psi}
\end{equation}
where the $\alpha_n$ are components of a normalized vector in $\Cx^{N+1}$. 
To calculate $\bra\Psi X\ket\Psi$, we need to know the $\bra mX\ket n$. We are 
especially interested in 
\begin{equation*}
 X = A \in\c M_2 
 \qquad\text{and}\qquad
 X = M \in \c M_2\otimes\c M_2.                                         
\end{equation*}
A possible trick is to consider 
\begin{equation*}
 X = \otimes_N \r e^{s\,A} = \idty + \sum_{j=1}^N A_j + \frac{s^2}{2}\, 
 \Bigl( \sum_{\{i,j \mid i\neq j\}} A_i\otimes A_j + \sum_{j=1}^N A_j^2 \Bigr) + \cdots
\end{equation*} 
with $A\in\c M_2$. Then 
\begin{align*}
 &\frac{\r d\r e^{s\, A}}{\r ds}\Bigr|_{s=0} = \sum_{j=1}^N A_j \\
 &\frac{\r d^2 \r e^{s\, A}}{\r ds^2}\Bigr|_{s=0} = \sum_{\{i,j \mid i\neq j\}} 
 A_i\otimes A_j + \sum_{i=1}^N A_i^2
\end{align*} 
and, because of symmetry,
\begin{align}
 &\bra mA\ket n = \frac{1}{N}\, \Bigl\langle m \,\Bigm|\, \frac{\r d\r
 e^{s\,A}}{\r ds} \Bigr|_{s=0} \,\Bigm|\, n \Bigr\rangle 
\label{eerste} \\
 &\bra mA\otimes A \ket n = \frac{1}{N(N-1)}\, \biggl( 
 \Bigl\langle m \,\Bigm|\, \frac{\r d^2\r e^{s\r A}}{\r ds^2}\Bigr|_{s=0}
 \,\Bigm|\, n \Bigr\rangle- \Bigl\langle m \,\Bigm|\, \sum_{j=1}^N A_j^2
 \,\Bigm|\, n \Bigr\rangle \biggr).
\label{tensor}
\end{align}
Now we have the following result
\begin{align*} 
 &\bigl\langle m \,\bigm|\, \otimes_N \r e^{s\,A} \,\bigm|\, n \bigr\rangle =
 \begin{pmatrix} N \\m \end{pmatrix}^{-1/2}
 \begin{pmatrix} N \\n \end{pmatrix}^{-1/2} \quad\times 
 \sum_{\uu=0}^{\max(n,m)} \\
 &\qquad\begin{pmatrix} N \\ \uu\,\ud\,\du\,\dd \end{pmatrix}
 \bigl( (\r e^{s\,A})_{\uparrow\uparrow} \bigr)^\uu\,
 \bigl( (\r e^{s\,A})_{\uparrow\downarrow} \bigr)^\ud\,
 \bigl( (\r e^{s\,A})_{\downarrow\uparrow} \bigr)^\du\,
 \bigl( (\r e^{s\,A})_{\downarrow\downarrow} \bigr)^\dd,
\end{align*}
with $m = \uu + \ud$ and $n = \uu + \du$ and $N = \uu + \ud + \du + \dd$. 
As seen in (\ref{eerste}), we can calculate the derivate of the previous formula 
and divide by $N$ to obtain $\bra nA\ket m$. This yields
\begin{align*}
 \bra mA \ket n 
 &= \frac{1}{N}\, \Bigl( m\, \delta_{m,n}\, A_{\uparrow\uparrow} +
 \sqrt{m(N-m+1)}\, \delta_{m,n-1}\, A_{\uparrow\downarrow} \\
 &\quad+ \sqrt{(m-1)(N-m)}\, \delta_{m-1,n}\, A_{\downarrow\uparrow} + (N-m)\,
 \delta_{m,n}\, A_{\downarrow\downarrow} \Bigr).
\end{align*} 
Similar computations with the second derivatives yield
\begin{align*}
 &\bra m B\otimes C \ket n
 = \frac{1}{N(N-1)} \Bigl[ 
 m(m-1)\, B_{\uparrow\uparrow} C_{\uparrow\uparrow}\, \delta_{m,n} \\
 &\quad+ m\, \sqrt{m(N-m+1)}\, \Bigl( B_{\uparrow\uparrow} C_{\uparrow\downarrow} 
 + B_{\uparrow\downarrow} C_{\uparrow\uparrow} \Bigr) \delta_{m-1,n} \\
 &\quad+ m\, \sqrt{(N-m)(m+1)}\, \Bigl( B_{\uparrow\uparrow} C_{\downarrow\uparrow}
 + B_{\downarrow\uparrow} C_{\uparrow\uparrow} \Bigr) \delta_{m+1,n} \\
 &\quad+ m\, (N-m)\, \Bigl( B_{\uparrow\uparrow} C_{\downarrow\downarrow}
 + B_{\downarrow\downarrow} C_{\uparrow\uparrow} \Bigr) \delta_{m,n} \\
 &\quad+ \sqrt{m(m-1)(N-m+2)(N-m+1)}\, B_{\uparrow\downarrow}
 C_{\uparrow\downarrow}\, \delta_{m-2,n} \\
 &\quad+ m\, (N-m)\, \Bigl( B_{\uparrow\downarrow} C_{\downarrow\uparrow}
 + B_{\downarrow\uparrow} C_{\uparrow\downarrow} \Bigr) \delta_{m,n} \\
 &\quad+ (N-m)\, \sqrt{m(N-m+1)}\, \Bigl( B_{\uparrow\downarrow} C_{\downarrow\downarrow}
 + B_{\downarrow\downarrow} C_{\uparrow\downarrow} \Bigr) \delta_{m-1,n} \\
 &\quad+ \sqrt{(m+2)(m+1)(N-m)(N-m-1)}\, B_{\downarrow\uparrow} C_{\downarrow\uparrow}\,
 \delta_{m+2,n} \\
 &\quad+ (N-m-1)\, \sqrt{(m+1)(N-m)}\, \Bigl( B_{\downarrow\uparrow} C_{\downarrow\downarrow}
 + B_{\downarrow\downarrow} C_{\downarrow\uparrow} \Bigr) \delta_{m+1,n} \\
 &\quad+ (N-m)(N-m-1)\, B_{\downarrow\downarrow} C_{\downarrow\downarrow}\,
 \delta_{m,n} \Bigr].
\end{align*}
In particular, 
\begin{equation*}
 {\textstyle \tr_{N-2}} \ket n\bra n = \frac{1}{4}\Bigl(P_1+P_{-1}+P_i+P_{-i}\Bigr) +\r O\Bigl( \frac{1}{N} \Bigr)
\end{equation*} 
where $P_\epsilon$ denotes the projection on $\otimes_2\frac{1}{\sqrt N}\Bigl(\sqrt n\, \ket \uparrow+\epsilon\, \sqrt{N-n}\,\ket \downarrow\Bigr)$ for $\epsilon=1,\,-1,\,i$ or $-i$. We obtain that $\ket n\bra n$ is separable up to a correction of order $\frac{1}{N}$. A similar computation shows that the marginal determined
by~(\ref{Psi}) is, up to order $1/N$, separable. The following theorem
provides a non-combinatorial answer to the question.

\begin{theo}
\label{approx}
 The distance between the two-site marginals of symmetric states on $N$ sites and
 the exchangeable two-site states is not larger than $d(d+1)/N$ where $d$ is the
 dimension of the single-site algebra. 
\end{theo}

\begin{proof}
Let us denote, for $B \in \c M_d^{\r h}(\Cx)$, by $B_j$ a copy of $B$ at site $j$.
By positivity and symmetry of an extension $\omega_N$ of $\omega$ we have
\begin{equation*}
 0 \le \omega_N\Bigl( \Bigl(\sum_{j=1}^N B_j\Bigr)^2 \Bigr) 
 = N(N-1) \omega(B \otimes B) + N \omega(B^2).   
\end{equation*}
Let $P^{\r s}$ be the projector on the symmetric subspace of $\Cx^d \otimes \Cx^d$,
which has dimension $d(d+1)/2$, then 
\begin{equation*}
 \tr P^{\r s}\, B\otimes B = \frac{1}{2}\, \tr B^2 
 + \frac{1}{2}\, \Bigl(\tr B \Bigr)^2\quad \text{for every } B\in\c M_d^{\r h}(\Cx). 
\end{equation*}
Choose now $c = Nd(d+1)/(N-1+d(d+1)) \le d(d+1)$ for $d=2,3,\ldots$ and $N=3,4,\ldots$, 
then for every $B\in\c M_d^{\r h}(\Cx)$
\begin{align*}
 &\Bigl( 1-\frac{c}{N} \Bigr)\, \omega(B\otimes B) + \frac{c}{N}\, \frac{2}{d(d+1)} 
 \tr P^{\r s} B\otimes B \\ 
 &\quad\ge - \Bigl( 1-\frac{c}{N} \Bigr)\, \frac{1}{N-1}\, \omega(B^2) + \frac{c}{Nd(d+1)} \tr B^2
 + \frac{c}{Nd(d+1)} \Bigl(\tr B\Bigr)^2 \\
 &\quad\ge \frac{1}{N-1+d(d+1)}\, \Bigl( \tr B^2 - \omega(B^2) \Bigr) \\
 &\quad\ge 0.
\end{align*}
Now by theorem~\ref{characterisation B-E} we get that 
\begin{equation*}
 X \in \c M_d(\Cx) \otimes \c M_d(\Cx) \mapsto \Bigl( 1-\frac{c}{N} \Bigr) \omega(X) + 
 \frac{c}{N}\, \frac{2}{d(d+1)} \tr P^{\r s} X
\end{equation*} 
is an exchangeable state. And so we have that
\begin{equation*}
 \r d(\c S^N,\c S^\infty) \le \frac{c}{N} \le \frac{d(d+1)}{N}.
\end{equation*}
\end{proof}

\section{Mean-field models of composite particles}

The Hamiltonian of a mean-field system of $N$ quantum spins with a pair
interaction $h$ is
\begin{equation}
 H^N = -\frac{2}{N} \sum_{\{i,j \mid 1\le i<j\le N\}} h_{ij}.
\label{mfh}
\end{equation} 
Here $h$ is a Hermitian matrix on $\Cx^d \otimes \Cx^d$ which is invariant
under the flip operation
\begin{equation*}
 \bra{\zeta\otimes\eta} h\,\ket{\varphi\otimes\psi} = 
 \bra{\eta\otimes\zeta} h\, \ket{\psi\otimes\varphi},\quad
 \eta,\zeta,\varphi,\psi \in \Cx^d.
\end{equation*}
We shall, moreover, assume that $h$ is ferromagnetic in the sense that there
exist $X^\alpha = \bigl(X^\alpha\bigr)^* \in \c M_d(\Cx)$ such that
\begin{equation}
 h = \sum_\alpha X^\alpha \otimes X^\alpha.
\label{ferro}
\end{equation}
The factor $2/N$ in~(\ref{mfh}) is needed to obtain a good thermodynamic
behaviour. 

A common example of such a model is the BCS-model where
\begin{align*}
 H^N
 &= - h\Bigl(\sum_{i=1}^N S_i^z\Bigr) - \frac{\lambda}{2\,N} 
 \Bigl(\sum_{i=1}^N S_i^+\Bigr) \Bigl(\sum_{j=1}^N S_j^-\Bigr) \\
 &= - h\Bigl(\sum_{i=1}^N S_i^z\Bigr) - \frac{\lambda}{2\,N} 
 \sum_{\{i,j=1 \mid i \neq j\}}^N \bigl( S_i^x S_j^x + S_i^y S_j^y \bigr) 
 + \r O(1). 
\end{align*}
Here $S^x$, $S^y$ and $S^z$ denote the generators of SU(2)
\begin{equation*}
 S^x = \frac{1}{2} \begin{pmatrix} 0 &1 \\1 &0 \end{pmatrix},\quad
 S^y = \frac{1}{2} \begin{pmatrix} 0 &-i\\i &0 \end{pmatrix},\quad
 S^z = \frac{1}{2} \begin{pmatrix} 1 &0\\0 &-1 \end{pmatrix},
\end{equation*}
and $S^\pm=S^x\pm\,i\,S^y$.

Using~(\ref{ferro}), we can rewrite the $N$-particle Hamiltonian
\begin{equation}
 H^N = - N \sum_\alpha \Bigl( \frac{1}{N} \sum_{i=1}^N X_i^\alpha \Bigr)^2 
 + \frac{1}{N} \sum_\alpha \Bigl( \sum_{i=1}^N \bigl( X_i^\alpha \bigr)^2
 \Bigr).
\label{neg}
\end{equation}
The second term in this expression has a norm of order 1 and is therefore
thermodynamically irrelevant. Therefore, up to a correction of order 1, $H_N$
is a sum of negative terms. Moreover, the average ground state 
energy can essentially be computed by varying over the fully symmetric pure states, 
which is a proper subclass of the symmetric states, sometimes called the 
\co{Bose symmetric states}.   

As with exchangeable states, there is the notion of \co{Bose exchangeable}
states. A state $\omega$ on $\otimes_N \c M_d(\Cx)$ is called Bose exchangeable 
if it admits for any $M>0$ a Bose symmetric extension $\omega_{(N+M)}$ to 
$\otimes_{N+M} \c M_d(\Cx)$. I.e., for any permutation $\pi$ of a set of 
$N+M$ points and any $A \in \otimes_{N+M} \c M_d(\Cx)$
\begin{equation}
 \omega_{(N+M)}(A) = \omega_{(N+M)}(A\, U_\pi)
\label{bex}
\end{equation} 
with $U_\pi$ as in~(\ref{rep}). Note that the asymmetry in
condition~(\ref{bex}) is only apparent as
\begin{equation*}
 \omega_{(N+M)}(U_\pi A) = \overline{\omega_{(N+M)}(A^* U_\pi)} 
 = \overline{\omega_{(N+M)}(A^*)} = \omega_{(N+M)}(A).
\end{equation*} 
The analogue of theorem~\ref{De Finetti} is then \cite{Hud}

\begin{theo}
\label{Hudson}
 If $\omega$ is a Bose exchangeable state on $\otimes_N \c M_d(\Cx)$, then
 \begin{equation*}
  \omega = \int_{\Cx^d_{\r{proj}}} \!\r d\mu([\varphi])\, \otimes_N [\varphi]
 \end{equation*}
 where $\Cx^d_{\r{proj}}$ is the complex projective $d$-dimensional Hilbert
 space and $\mu$ is a probability measure on $\Cx^d_{\r{proj}}$. By $[\varphi]$ 
 we denote the pure state of $\c M_d(\Cx)$ determined by the subspace 
 $\Cx\varphi$ with $\norm\varphi = 1$, i.e.\
 \begin{equation*}
  [\varphi](A) := \bra\varphi A\ket{\varphi},\quad A\in\c M_d(\Cx). 
 \end{equation*} 
\end{theo}

The asymptotic ground state energy density of a mean-field Hamiltonian with
pair interaction $h$ is then given by
\begin{equation*}
 e_0(h) := \lim_{N\to\infty} \frac{1}{N}\,  \inf_\omega \omega(H_N).
\end{equation*}
Because of the permutation invariance and of condition~(\ref{ferro}), we
have
\begin{equation}
 e_0(h) =- \max_{[\varphi]}\ \Bigl( [\varphi]\otimes[\varphi](h) \Bigr). 
\label{gs}
\end{equation}  
Indeed, by theorem~\ref{De Finetti} it suffices to compute the infimum over
product exchangeable states and if 
\begin{equation*}
 \rho = \sum_i r_i \ket{\varphi_i} \bra{\varphi_i}
\end{equation*} 
is the eigenvalue decomposition of $\rho$ we have, using
condition~(\ref{ferro}) and the convexity of $x \mapsto x^2$
\begin{align*}
 \rho\otimes\rho(h) 
 &= \sum_\alpha \rho(X_\alpha)^2 = \sum_\alpha 
 \Bigl( \sum_i r_i\, [\varphi_i](X_\alpha) \Bigr)^2 \\
 &\le \sum_\alpha \sum_i r_i \bigl( [\varphi_i](X_\alpha) \bigr)^2
 = \sum_i r_i\, [\varphi_i]\otimes[\varphi_i](h).
\end{align*}

The state space of a composite particle is of the form $\Cx^{d_1} \otimes
\Cx^{d_2}$. We shall consider a simple pair interaction $h_{12} = h_1
\otimes h_2$ between such pair interactions with $h_1$ and $h_2$
ferromagnetic in the sense of~(\ref{ferro}). We now have the following result

\begin{theo}
\label{gsm}
 Assume that $h_i \in \c M_{d_i}(\Cx) \otimes \c M_{d_i}(\Cx)$, i=1,2 are
 Hermitian, invariant under the flip and satisfy condition~(\ref{ferro}).
 Assume, moreover, that $h_1$ is positive definite, then
 \begin{equation*}
  e_0(h_1 \otimes h_2) = - e_0(h_1)\, e_0(h_2).
 \end{equation*}  
\end{theo}

\begin{proof}
By the negativity of the mean-field Hamiltonians corresponding to
pair-interactions satisfying~(\ref{ferro}), see~(\ref{neg}), we have
\begin{align*}
 e_0(h_1 \otimes h_2) 
 &= -\max_{[\varphi_{12}]}\ \Bigl( [\varphi_{12}] \otimes [\varphi_{12}](h_1\otimes h_2) \Bigr) \\
 &\le - \max_{\{[\varphi_{12}] \mid [\varphi_{12}] = [\varphi_1] \otimes [\varphi_2]\}}\ 
 \Bigl( [\varphi_{12}] \otimes [\varphi_{12}](h_1\otimes h_2) \Bigr) \\
 &= - \max_{[\varphi_1]}\ \Bigl( [\varphi_1]\otimes[\varphi_1](h_1) \Bigr)\ 
 \max_{[\varphi_2]}\ \Bigl( [\varphi_2]\otimes[\varphi_2](h_2) \Bigr) \\
 &= - e_0(h_1)\, e_0(h_2).
\end{align*}

To obtain the converse inequality, consider a normalized vector $\varphi_{12} \in
\Cx^{d_1} \otimes \Cx^{d_2}$ and the state
\begin{equation*}
 \omega_2^{[\varphi_{12}]}(x) := \frac{[\varphi_{12}] \otimes
 [\varphi_{12}] (h_1 \otimes x)}{[\varphi_{12}] \otimes
 [\varphi_{12}] (h_1 \otimes \idty)}
\end{equation*} 
on $\c M_{d_2}(\Cx) \otimes \c M_{d_2}(\Cx)$. This state is flip-invariant and,
because 
\begin{equation*}
 h_1 = \sum_\alpha X^\alpha \otimes X^\alpha
\end{equation*}
enjoys the property
\begin{equation*}
 \omega_2^{[\varphi_{12}]}(Y \otimes Y) \ge 0,\quad Y = Y^* \in \c M_{d_2}(\Cx). 
\end{equation*}
Hence, by theorem~\ref{characterisation B-E}, it is a mixture of product
states. Then by the remarks above
\begin{equation*}
 \omega_2^{[\varphi_{12}]}(h_2) \ge e_0(h_2).
\end{equation*}
We therefore have
\begin{align*}
 e_0(h_1 \otimes h_2) 
 &= -\max_{[\varphi_{12}]}\ \Bigl( [\varphi_{12}] \otimes [\varphi_{12}](h_1\otimes h_2) \Bigr) \\
 &= -\max_{[\varphi_{12}]}\ \Bigl( [\varphi_{12}] \otimes [\varphi_{12}] (h_1
 \otimes \idty)\ \omega_2^{[\varphi_{12}]} (h_2) \Bigr) \\
 &\ge e_0(h_2) \max_{[\varphi_{12}]}\ \Bigl( [\varphi_{12}] \otimes [\varphi_{12}] (h_1
 \otimes \idty) \Bigr) \\
 &\ge - e_0(h_1)\, e_0(h_2).
\end{align*}
The last estimate follows from the fact that $\idty$ is positive definite
and satisfies condition~(\ref{ferro}).
\end{proof}

Two remarks are here in order. There doesn't seem to be a simple extension
of theorem~\ref{gsm} to finite temperatures, at least no simple relation
between the free energy densities of the composite system and the components
seems to exist. A second remark is that the theorem can be used to give a
partial answer to the problem of multiplicativity of maximal 2-norm of
quantum channels~\cite{Fuk,Kin}. Unfortunately, the positivity 
condition on $h_1$ imposes some restriction on the allowed channels. A 
further elaboration of this matter will be considered in a future publication.  

\section*{Appendix A: The map $\bi{V_d}$}

Every Hermitian matrix $B$ in $\c M^{\r h}_d(\Cx)$ can be written as
\begin{equation*}
 B = \begin{pmatrix}
      b&\bra{\psi}\\
      \ket{\psi}& B_0
     \end{pmatrix}
\end{equation*} 
where $b\in\Rl$, $\ket{\psi}$ is a vector in $\Cx^{d-1}$ and $B_0$ a matrix
in $\c M^{\r h}_{d-1}(\Cx)$. We then define the map $V_d: \c M^{\r h}_d(\Cx) 
\rightarrow \Rl^{d^2}$ inductively as
\begin{equation*}
 V_d(B) := \begin{pmatrix}
            b \\
            \sqrt2\, \R\ket\psi \\
            \sqrt2\,\I\ket\psi \\
            V_{d-1}(B_0)
           \end{pmatrix}.
\end{equation*}
This map has the following properties for $B_1,\,B_2 \in \c M^{\r h}_d(\Cx)$
\begin{itemize}
\item[i)] 
 $V_d(B_1 + B_2) = V_d(B_1) + V_d(B_2)$.
\item[ii)] 
 For every $\lambda\in\Rl$, $V_d(\lambda\,B_1) = \lambda V_d(B_1)$.
\item[iii)] 
 $\tr B_1\,B_2 = \bra{V_d(B_1)}V_d(B_2)\rangle$. 
\end{itemize}
This can easily be proved by induction on $d$. Moreover, the map $V_d$ is 
one-to-one and onto. Note, however, that the map $V_d$ is basis dependent.

\section*{Appendix B: The map $\bi{M_d}$}

\subsubsection*{The subspace $\bi{\c K}$}

Before we start to search for a good map $M_d$, we take a closer look at the 
subset $\c K$ of flip-symmetric, complex, hermitian matrices on $\Cx^{d^2}$. We begin 
by decomposing the $d$-dimensional Hilbert space $\Cx^d$ in a direct sum of
a one-dimensional and a $(d-1)$-dimensional space, $\Cx^d = \Cx \oplus \Cx^{d-1}$. 
We are interested in the symmetric, $(\Cx^d \otimes \Cx^d)^{\r s}$, and 
antisymmetric, $(\Cx^d \otimes \Cx^d)^{\r a}$, subspaces of 
$\Cx^d \otimes \Cx^d$ as they are the ones left invariant by the elements in $C^*$.  
We consider a basis $\{ e_0,\ldots,e_{d-1}\}$ of $\Cx^d$. Then a basis of 
$(\Cx^d \otimes \Cx^d)^{\r s}$ is given by
\begin{equation*}
 \{e_0\otimes e_0, g_1, \ldots, g_{d-1}, f_1, \dots, f_{d(d-1)/2} \}
\label{basissym}
\end{equation*}
where $g_i := \frac{1}{\sqrt2} (e_0\otimes e_i + e_i\otimes e_0)$ and where 
the $f_i$ generate the symmetric subspace of $\Cx^{d-1} \otimes \Cx^{d-1}$. 
Similarly, a basis of $(\Cx^d \otimes \Cx^d)^{\r a}$ is given by
\begin{equation*}
 \{h_1, \ldots, h_{d-1}, k_1, \ldots, k_{(d-2)(d-1)/2} \}
\label{basisanti}
\end{equation*}
where $h_i := \frac{1}{\sqrt2} (e_0\otimes e_i - e_i\otimes e_0)$ and where 
the $k_i$ generate the antisymmetric subspace of $\Cx^{d-1} \otimes \Cx^{d-1}$.

A matrix $A\in\c K$ can be written in this symmetric-antisymmetric basis as
\begin{equation}
 A = \begin{pmatrix}
      a &\bra\varphi &\bra\Phi &0 &0 \\
      \ket\varphi &X_1 &Y_1 &0 &0 \\
      \ket\Phi &Y_1^* &Z_1 &0 &0 \\
      0 &0 &0 &X_2 &Y_2 \\
      0 &0 &0 &Y_2^* &Z_2
     \end{pmatrix}
\label{star}
\end{equation}
where
\begin{align*}
 &a\in\Cx,\ \varphi\in\Cx^{d-1},\ \Phi\in \bigl( \Cx^{d-1} \otimes 
 \Cx^{d-1} \bigr)^{\r s},\ 
 X_1,X_2 \in \c M_{d-1}(\Cx) \\
 &Z_1: \bigl( \Cx^{d-1} \otimes \Cx^{d-1} \bigr)^{\r s} \rightarrow 
 \bigl( \Cx^{d-1} \otimes \Cx^{d-1} \bigr)^{\r s},\ 
 Z_2: \bigl( \Cx^{d-1} \otimes \Cx^{d-1} \bigr)^{\r a} \rightarrow
 \bigl( \Cx^{d-1} \otimes \Cx^{d-1} \bigr)^{\r a} \\
 &Y_1: \bigl( \Cx^{d-1} \otimes \Cx^{d-1} \bigr)^{\r s} \rightarrow
 \Cx^{d-1}
 \qquad\text{and}\qquad
 Y_2: \bigl( \Cx^{d-1} \otimes \Cx^{d-1} \bigr)^{\r a} \rightarrow
 \Cx^{d-1}.
\end{align*}

In order to ensure that we map the subspace $\c K$ in a suitable vector space, 
we can count its real dimension. The restriction of elements of $\c K$
to the symmetric subspace needs $d(d+1)/2$ real parameters on the diagonal and 
2 times (for the real and imaginary parts) $[d(d+1)/2] [(d(d+1)/2) - 1]/2$ off 
the diagonal. For the restriction to the antisymmetric subspace we need $d(d-1)/2) + 
[d(d-1)/2][(d(d-1)/2) - 1 ]$ parameters. In total this amounts to
$d^2(d^2 + 1)/2$ real parameters, which is exactly equal to the dimension of the 
symmetric real matrices of dimension $d^2$, i.e.\ the matrices $M \in \c M_{d^2}(\Rl)$
such that $M = M^{\s T}$ where ${\s T}$ denotes transposition.
 
\subsubsection*{The map $\bi{M_d}$}

Denote the symmetric real matrices of dimension $d^2$ by $\c M^{\r h}_{d^2}(\Rl)$.
Using the parametrisation~(\ref{star}) for $A\in\c K$ we define the map 
$M_d: \c K \rightarrow \c M^{\r h}_{d^2}(\Rl)$ by
\begin{equation*}
 M_d(A) := \mbox{\scriptsize
           \mbox{$ 
           \begin{pmatrix}
            a 
            &\bra{\R\varphi}
            &\bra{\I\varphi} 
            &\Bigm| V_{d-1} \Bigl( {\displaystyle\frac{X_1+X_2}{2}} \Bigr) 
            \Bigr\rangle \\[3pt]
            \ket{\R\varphi}
            &{\displaystyle \frac{\R X_1 - \R X_2}{2}} + [\R\Phi]
            &{\displaystyle \frac{\I X_1 - \I X_2}{2}} + [\I\Phi]
            &T_1(Y_1,Y_2) \\[3pt]
            \ket{\I\varphi}
            &\Bigl( {\displaystyle\frac{\I X_1 - \I X_2}{2}} + [\I\Phi] \Bigr)^{\s T}
            &{\displaystyle\frac{\R X_1 - \R X_2}{2}} - [\R\Phi]
            &T_2(Y_1,Y_2) \\[3pt]
            \Bigl\langle V_{d-1} \Bigl( {\displaystyle \frac{X_1+X_2}{2}} \Bigr) \Bigm|
            &T_1(Y_1,Y_2)^{\s T}
            &T_2(Y_1,Y_2)^{\s T}
            &M_{d-1} \begin{pmatrix}
                Z_1&0 \\
                0&Z_2
               \end{pmatrix}
           \end{pmatrix}
           $}}
\end{equation*} 
where for $i \neq j$
\begin{equation*}
 [\R\Phi]_{ii} := \R\Phi_{ii},\   
 [\R\Phi]_{ij} := \frac{1}{\sqrt2} \R\Phi_{ij},\ 
 [\I\Phi]_{ii} := \I\Phi_{ii}\ \text{and } 
 [\I\Phi]_{ij} := \frac{1}{\sqrt2} \I\Phi_{ij}.
\end{equation*}
We describe the maps $T_1$ and $T_2$ in the two following paragraphs.
As with $V_d$, the map $M_d$ is basis dependent

\paragraph{The map $\bi{T_1}$}

Recalling that $\{e_i\}_{i=1}^{d-1}$ is a basis we choose in $\Cx^{d-1}$, 
let us, for $i<j$, $i,j=1,\ldots,d-1$ and any matrix $B_0 \in \c M^{\r h}_{d-1}(\Cx)$
put
\begin{align*}
 &\beta_R(i,j) := \alpha\qquad \text{if and only if } 
 \bra{V_{d-1}(B_0)} e_\alpha\rangle = \sqrt2\, \R[B_0]_{ij} \\
 &\beta_I(i,j) := \alpha\qquad \text{if and only if } 
 \bra{V_{d-1}(B_0)} e_\alpha\rangle = \sqrt2\, \I[B_0]_{ij}\qquad \text{and} \\
 &\beta(i) := \alpha\qquad \text{if and only if } 
 \bra{V_{d-1}(B_0)} e_\alpha\rangle = [B_0]_{ii}.
\end{align*}
This way of denoting the matrix elements will be useful later on when we 
will compare $B\otimes B$ with the projection on $V_d(B)$. We also define 
$\epsilon_k^\ell=1$ if $k<\ell$ and $-1$ otherwise.
We are now ready to define the map $T_1$ by looking at each of the matrix elements. 
In the following, $i,k,\ell$ run from $1$ to $d-1$ and $i<\ell$, $i\neq k$, $\ell\neq k$
\begin{itemize}
\item
 $[T_1(Y_1,Y_2)]_{i,\beta(i)} := \R[Y_1]_{i,ii}$
\item
 $[T_1(Y_1,Y_2)]_{k,\beta(i)} := \frac{1}{\sqrt2}\, \R\Bigl( [Y_1]_{i,ik} 
 + \epsilon_k^i\, [Y_2]_{i,ik} \Bigr)$
\item
 $[T_1(Y_1,Y_2)]_{i,\beta(i,\ell)} := \frac{1}{\sqrt2}\, \R\Bigl( [Y_1]_{\ell,ii}
 + {\displaystyle \frac{[Y_1]_{i,i\ell} + \epsilon_i^\ell\, [Y_2]_{i,i\ell}}{\sqrt2}} 
 \Bigr)$
\item
 $[T_1(Y_1,Y_2)]_{\ell,\beta_R(i,\ell)} := \frac{1}{\sqrt2}\, \R\Bigl( [Y_1]_{i,\ell\ell}
 + {\displaystyle \frac{[Y_1]_{\ell,i\ell} + \epsilon_\ell^i\, [Y_2]_{\ell,i\ell}}{\sqrt2}}
 \Bigr)$
\item
 $[T_1(Y_1,Y_2)]_{k,\beta_R(i,\ell)} := \R\Bigl( {\displaystyle \frac{[Y_1]_{\ell,ik}
 + \epsilon_k^i\, [Y_2]_{\ell,ik} + [Y_1]_{i,\ell k} 
 + \epsilon_k^\ell\, [Y_2]_{i,\ell k}}{2}} \Bigr)$
\item
 $[T_1(Y_1,Y_2)]_{i,\beta_I(i,\ell)} := -\frac{1}{\sqrt2}\, \I\Bigl( [Y_1]_{\ell,ii}
 - {\displaystyle \frac{[Y_1]_{i,i\ell} + \epsilon_i^\ell\, [Y_2]_{i,i\ell}}{\sqrt2}} 
 \Bigr)$
\item
 $[T_1(Y_1,Y_2)]_{\ell,\beta_I(i,\ell)} := \frac{1}{\sqrt2}\, \I\Bigl( [Y_1]_{i,\ell\ell}
 - {\displaystyle \frac{[Y_1]_{\ell,i\ell} + \epsilon_\ell^i\, [Y_2]_{\ell,i\ell}}{\sqrt2}}
 \Bigr)$
\item
 $[T_1(Y_1,Y_2)]_{k,\beta_I(i,\ell)} := -\I\Bigl( {\displaystyle \frac{[Y_1]_{\ell,ik}
 + \epsilon_k^i\,[Y_2]_{\ell,ik} - [Y_1]_{i,\ell k} 
 - \epsilon_k^\ell\, [Y_2]_{i,\ell k}}{2}} \Bigr)$
\end{itemize}

\paragraph{The map $\bi{T_2}$}

The notations are similar to the ones used for the map $T_1$. Again we define 
each matrix element
\begin{itemize}
\item
 $[T_2(Y_1,Y_2)]_{i,\beta(i)} := \I[Y_1]_{i,ii}$
\item
 $[T_2(Y_1,Y_2)]_{k,\beta(i)} := \frac{1}{\sqrt2}\, \I\Bigl( [Y_1]_{i,ik} 
 + \epsilon_k^i\, [Y_2]_{i,ik} \Bigr)$
\item
 $[T_2(Y_1,Y_2)]_{i,\beta_R(i,\ell)} := \frac{1}{\sqrt2}\, \I\Bigl( [Y_1]_{\ell,ii}
 + {\displaystyle \frac{[Y_1]_{i,i\ell} + \epsilon_i^\ell\, [Y_2]_{i,i\ell}}{\sqrt2}}
 \Bigr)$
\item
 $[T_2(Y_1,Y_2)]_{\ell,\beta_R(i,\ell)} := \frac{1}{\sqrt2}\, \I\Bigl( [Y_1]_{i,\ell\ell}
 + {\displaystyle \frac{[Y_1]_{\ell,i\ell} + \epsilon_\ell^i\, [Y_2]_{\ell,i\ell}}{\sqrt2}}
 \Bigr)$
\item
 $[T_2(Y_1,Y_2)]_{k,\beta_R(i,\ell)} := \I\Bigl( {\displaystyle \frac{[Y_1]_{\ell,ik} 
 + \epsilon_k^i\, [Y_2]_{\ell,ik} + [Y_1]_{i,\ell k} 
 + \epsilon_k^\ell\, [Y_2]_{i,\ell k}}{2}} \Bigr)$
\item
 $[T_2(Y_1,Y_2)]_{i,\beta_I(i,\ell)} := \frac{1}{\sqrt2}\, \R\Bigl( [Y_1]_{\ell,ii}
 - {\displaystyle \frac{[Y_1]_{i,i\ell} + \epsilon_i^\ell\, [Y_2]_{i,i\ell}}{\sqrt2}}
 \Bigr)$
\item
 $[T_2(Y_1,Y_2)]_{\ell,\beta_I(i,\ell)} := -\frac{1}{\sqrt2} \R\Bigl( [Y_1]_{i,\ell\ell}
 - {\displaystyle \frac{[Y_1]_{\ell,i\ell} + \epsilon_\ell^i\, [Y_2]_{\ell,i\ell}}{\sqrt2}}
 \Bigr)$
\item
 $[T_2(Y_1,Y_2)]_{k,\beta_I(i,\ell)} := \R\Bigl( {\displaystyle \frac{[Y_1]_{\ell,ik} 
 + \epsilon_k^i\, [Y_2]_{\ell,ik} - [Y_1]_{i,\ell k} 
 - \epsilon_k^\ell\, [Y_2]_{i,\ell k}}{2}} \Bigr)$.
\end{itemize}
One can easily see that, given $T_1(Y_1,Y_2)$ and $T_2(Y_1, Y_2)$, one can 
reconstruct the matrices $Y_1$ and $Y_2$. Also these two maps are real linear.

\subsubsection*{Properties of the map $\bi{M_d}$}

The map $M_d$ has similar properties as the map $V_d$
\begin{itemize}
\item 
 $M_d(A_1+A_2) = M_d(A_1) + M_d(A_2)$.
\item 
 For every $\lambda\in\Rl$, $M_d(\lambda\,A) = \lambda\,M_d(A)$.
\end{itemize}
It is also one-to-one and onto. Again one can easily check these properties 
by induction on $d$ using $\I(X_1^{ij} - X_2^{ij}) = -\I(X_1^{ji} - X_2^{ji})$.

\subsubsection*{The image of $\bi{B\otimes B}$}

Fix $B\in\c M^{\r h}_d(\Cx)$ and consider the tensor product of $B$ with itself 
\begin{equation*}
 B\otimes B = \begin{pmatrix}
               b^2
               &\sqrt2\, b\bra\psi
               &\bra{(\psi\otimes\psi)^{\r s}}
               &0
               &0 \\
               \sqrt2\, b\ket\psi
               &b\, B_0 + \ket\psi\bra\psi
               &(\bra\psi \otimes B_0)^{\r s}
               &0
               &0 \\
               \ket{(\psi\otimes\psi)^{\r s}}
               &(\ket\psi \otimes B_0)^{\r s}
               &(B_0\otimes B_0)^{\r s}
               &0
               &0 \\
               0
               &0
               &0
               &b \,B_0 - \ket\psi\bra\psi
               &(\bra\psi \otimes B_0)^{\r a} \\
               0
               &0
               &0
               &(\ket\psi \otimes B_0)^{\r a}
               &(B_0\otimes B_0)^{\r a}
              \end{pmatrix}
\end{equation*} 
We will prove that $B\otimes B$ is mapped by $M_d$ on $\ket{V_d(B)} \bra{V_d(B)}$ 
with
\begin{equation*}
 V_d(B) = \begin{pmatrix} 
           b \\
           \sqrt2 \bra{\R\psi} \\
           \sqrt2 \bra{\I\psi}\\
	    V_{d-1}(B_0)
          \end{pmatrix}.
\end{equation*} 
First we write down the image of $B\otimes B$                                                                                            
\begin{align*}
 &M_d(B\otimes B) \\
 &=\mbox{\scriptsize \mbox{$\begin{pmatrix}
  b^2
  &b\, \sqrt2 \bra{\R\psi}
  &b\, \sqrt2 \bra{\I\psi}
  &b\, \bra{V_{d-1}(B_0)} \\
  b\, \sqrt2 \ket{\R\psi}
  &\R\ket\psi\bra\psi + [\R (\psi\otimes\psi)^{\r s}]
  &\I\ket\psi\bra\psi + [\I (\psi\otimes\psi)^{\r s}]
  &T_1((\bra\psi \otimes B_0)^{\r s}, (\bra\psi\otimes B_0)^{\r a}) \\
  b\, \sqrt2 \ket{\I\psi}
  &\I\ket\psi\bra\psi + [\I (\psi\otimes\psi)^{\r s}]^*
  &\R\ket\psi\bra\psi - [\R (\psi\otimes\psi)^{\r s}]
  &T_2((\bra\psi \otimes B_0)^{\r s}, (\bra\psi \otimes B_0)^{\r a}) \\
  b\, \ket{V_{d-1}(B_0)}
  &T_1((\bra\psi \otimes B_0)^{\r s}, (\bra\psi \otimes B_0)^{\r a})^*
  &T_2((\bra\psi \otimes B_0)^{\r s}, (\bra\psi \otimes B_0)^{\r a})^*
  &M_{d-1}(B_0\otimes B_0)
 \end{pmatrix}$}}
\end{align*}

The first row and column are encouraging but we still have some steps to verify. 
If we use induction on $d$, we also get that $M_{d-1}(B_0\otimes B_0) = 
\ket{V_{d-1}(B_0)} \bra{V_{d-1}(B_0)}$. Let's look at the other parts of the matrix.

Looking at the elements in the middle of the matrices $M_d(B\otimes B)$, we need to 
prove that 
\begin{itemize}
\item 
 $\R\ket\psi\bra\psi + [\R (\psi\otimes\psi)^{\r s}] = \ket{\sqrt2\, \R\psi}
 \bra{\sqrt2\, \R\psi}$,
\item 
 $\R\ket\psi\bra\psi - [\R (\psi\otimes\psi)^{\r s}] = \ket{\sqrt2\, \I\psi}
 \bra{\sqrt2\, \I\psi}$ and
\item 
 $\I\ket\psi\bra\psi + [\I (\psi\otimes\psi)^{\r s}] = \ket{\sqrt2\, \R\psi}
 \bra{\sqrt2\, \I\psi}$
\end{itemize}
in order to obtain that $B\otimes B$ is mapped on $\ket{V(B)}\bra{V(B)}$. 

Let's look at the different matrix elements
\begin{itemize}
\item 
 $\R\ket\psi\bra\psi + [\R (\psi\otimes\psi)^{\r s}] = 2\ket{\R\psi}\bra{\R\psi}$.
 Indeed, it is easy to see that 
 \begin{align*}
  &[\R\ket\psi\bra\psi + [\R (\psi\otimes\psi)^{\r s}]]_{ii} = 
  \bigl( (\R\psi_i)^2 + (\I\psi_i)^2 \bigr) + \R\psi_i^2
  = 2(\R\psi_i)^2 \qquad \text{and} \\
  &[\R\ket\psi\bra\psi + [\R (\psi\otimes\psi)^{\r s}]]_{ij}
  = \R\psi_i\, \R\psi_j + \I\psi_i\, \I\psi_j + \frac{1}{\sqrt2}\, 
  \R(\sqrt2\psi_i\,\psi_j) \\
  &= 2\,\R\psi_i\, \R\psi_j.
 \end{align*}
\item 
 $\R\ket\psi\bra\psi - [\R (\psi\otimes\psi)^{\r s}] = 2\,\ket{\I\psi}
 \bra{\I\psi}$. The proof is similar to the one above.
\item 
 $\I\ket\psi\bra\psi + [\I (\psi\otimes\psi)^{\r s}] = \ket{\R\psi}
 \bra{\I\psi}$. Indeed,
 \begin{align*}
  &[\I\ket\psi\bra\psi + [\I (\psi\otimes\psi)^{\r s}]]_{ii}
  =2\, \R\psi_i \I\psi_i \qquad \text{and} \\
  &[\I\ket\psi\bra\psi + [\I (\psi\otimes\psi)^{\r s}]]_{ij}
  =\R\psi_i\, \I\psi_j - \I\psi_i\, \R\psi_j + \frac{1}{\sqrt2}\, 
  \I(\sqrt2\,\psi_i\,\psi_j) \\
  &=2\, \R\psi_i\, \I\psi_j.
\end{align*}
\end{itemize}

Now the proof is almost complete. We still have to verify that that 
$(\bra\psi \otimes B_0)^{\r s}$ and $(\bra\psi \otimes B_0)^{\r a}$ are 
mapped by $T_1$ and $T_2$ on $\ket{\sqrt2\, \R\psi} \bra{V_{d-1}(B_0)}$ 
and $\ket{\sqrt2\, \I\psi} \bra{V_{d-1}(B_0)}$ respectively.

\paragraph{The map $\bi{T_1}$}

We now verify that 
\begin{equation*}
 T_1((\bra\psi \otimes B_0)^{\r s}, (\bra\psi \otimes B_0)^{\r a})
 = \ket{\sqrt2\, \R\psi} \bra{V_{d-1}(B_0)}.
\end{equation*}

\begin{align*}
\bullet\, 
 [T_1((\bra\psi \otimes B_0)^{\r s}, 
 &(\bra\psi\otimes B_0)^{\r a})]_{i,\beta(i)}
 = \R(\sqrt2\,\psi_i\,[B_0]_{ii}) = \sqrt2\, \R\psi_i\, [B_0]_{ii} \\
\bullet\,
 [T_1((\bra\psi \otimes B_0)^{\r s}, 
 &(\bra\psi \otimes B_0)^{\r a})]_{k,\beta(i)} \\
 &= \frac{1}{\sqrt2} \R(\psi_i\,[B_0]_{ik} + \psi_k\,[B_0]_ii 
 + \epsilon_k^i\, \epsilon_i^k (\psi_i\, [B_0]_{ik} - \psi_k\, [B_0]_{ii})) \\
 &= \sqrt2\, \R\psi_k\, [B_0]_{ii} \\
\bullet\,
 [T_1((\bra\psi \otimes B_0)^{\r s},
 &(\bra\psi \otimes B_0)^{\r a})]_{i,\beta_R(i,\ell)} \\
 &= \frac{1}{\sqrt2} \R(\sqrt2\psi_i\, [B_0]_{\ell i} + \frac{\psi_i\, [B_0]_{i\ell}
 + \psi_\ell\, [B_0]_{ii} + (\psi_i\, [B_0]_{i\ell} - \psi_\ell\, [B_0]_{ii})}
 {\sqrt2}) \\
 &= \R(\psi_i([B_0]_{i\ell} + [B_0]_{\ell i})) = \sqrt2\, \R\psi_i\,
 \sqrt2\, \R\,[B_0]_{i\ell} \\
\bullet\,
 [T_1((\bra\psi \otimes B_0)^{\r s},
 &(\bra\psi \otimes B_0)^{\r a})]_{\ell,\beta_R(i,\ell)} \\
 &= \frac{1}{\sqrt2} \R(\sqrt2 \psi_\ell\, [B_0]_{i\ell} + \frac{\psi_i\, [B_0]_{\ell\ell}
 + \psi_\ell\, [B_0]_{\ell i} - (\psi_i\, [B_0]_{\ell\ell} - \psi_\ell\, [B_0]_{\ell i})}
 {\sqrt2}) \\
 &= \R(\psi_\ell([B_0]_{i\ell} + [B_0]_{\ell i})) = \sqrt2\, \R\psi_\ell\,
 \sqrt2\, \R[B_0]_{i\ell} \\
\bullet\,
 [T_1((\bra\psi \otimes B_0)^{\r s},
 &(\bra\psi \otimes B_0)^{\r a})]_{k,\beta_R(i,\ell)} \\
 &= \frac{1}{2} \R(\psi_i\, [B_0]_{\ell k} + \psi_k\, [B_0]_{\ell i} 
 + \epsilon_k^i\, \epsilon_i^k (\psi_i\, [B_0]_{\ell k} - \psi_k\, [B_0]_{\ell i}) \\
 &\qquad\qquad + \psi_\ell\, [B_0]_{ik} + \psi_k\, [B_0]_{i\ell} + \epsilon_k^\ell\,
 \epsilon_\ell^k (\psi_\ell\, [B_0]_{ik} - \psi_k\, [B_0]_{i\ell})) \\
 &= \R\,\psi_k([B_0]_{i\ell} + [B_0]_{\ell i}) = \sqrt2\, \R\,\psi_k\,
 \sqrt2\, \R[B_0]_{i\ell} \\
\bullet\,
 [T_1((\bra\psi \otimes B_0)^{\r s},
 &(\bra\psi \otimes B_0)^{\r a})]_{i,\beta_I(i,\ell)} \\
 &= \frac{1}{\sqrt2} \I(-\sqrt2 \psi_i\, [B_0]_{\ell i} + \frac{\psi_i\, [B_0]_{i\ell}
 + \psi_\ell\, [B_0]_{ii} - (\psi_i\, [B_0]_{i\ell} - \psi_\ell\, [B_0]_{ii})}{\sqrt2}) \\
 &= \I(\psi_i([B_0]_{i\ell} - [B_0]_{\ell i})) = \sqrt2\, \R\psi_i\,
 \sqrt2\, \I[B_0]_{i\ell} \\
\bullet\,
 [T_1((\bra\psi \otimes B_0)^{\r s},
 &(\bra\psi \otimes B_0)^{\r a})]_{\ell,\beta_I(i,\ell)} \\
 &= \frac{1}{\sqrt2} \I(\sqrt2\psi_\ell\, [B_0]_{i\ell} - \frac{\psi_i\, [B_0]_{\ell\ell}
 +\psi_\ell\, [B_0]_{\ell i} - (\psi_i\, [B_0]_{\ell\ell} - \psi_\ell\, [B_0]_{\ell i})}
 {\sqrt2}) \\
 &= \I(\psi_l\ell([B_0]_{i\ell} - [B_0]_{\ell i})) = \sqrt2\, \R\psi_\ell\,
 \sqrt2\, \I[B_0]_{i\ell} \\
\bullet\,
 [T_1((\bra\psi \otimes B_0)^{\r s},
 &(\bra\psi \otimes B_0)^{\r a})]_{k,\beta_I(i,\ell)} \\
 &= \frac{1}{2} \I(-\psi_i\, [B_0]_{\ell k} - \psi_k\, [B_0]_{\ell i} 
 - \epsilon_k^i\, \epsilon_i^k (\psi_i\, [B_0]_{\ell k} - \psi_k\, [B_0]_{\ell i}) \\
 &\qquad\qquad + \psi_\ell\, [B_0]_{ik} + \psi_k\, [B_0]_{i\ell} + \epsilon_k^\ell\,
 \epsilon_\ell^k (\psi_\ell\, [B_0]_{ik} - \psi_k\, [B_0]_{i\ell})) \\
 &= \I\psi_k([B_0]_{i\ell} - [B_0]_{\ell i}) = \sqrt2\, \R\psi_k\,
 \sqrt2\, \I[B_0]_{i\ell}
\end{align*}

\paragraph{The map $\bi{T_2}$}

The proof that 
\begin{equation*}
 T_2((\bra\psi \otimes B_0)^{\r s}, (\bra\psi \otimes B_0)^{\r a}))
 = \ket{\sqrt2\, \I\psi} \bra{V_{d-1}(B_0)}
\end{equation*} 
is completely similar, so we will not provide the details. We have now proven a one-to-one 
correspondence between $B\otimes B \in \bigl( \c M_d(\Cx) \otimes \c M_d(\Cx) 
\bigr)^{\r h}$ and 
the subset of rank one projections in $\c M_{d^2}(\Rl)$. We now have real linear
one-to-one and onto maps $V_d$ and 
$M_d$ that satisfy condition~$(iii)$ Section~\ref{quantum case}. 
Let us now examine condition~$(ii)$.

\section*{Appendix C: $\bi{\tr A\,(B\otimes B) = \bra{V_d(B)} M_d(A) \ket{V_d(B)}}$}

We start by calculating the trace of $A(B\otimes B)$. 
\begin{align*}
 \tr A\, (B\otimes B)
 &=\tr A^{\r s}\, (B\otimes B)^{\r s} + \tr A^{\r a}\, (B\otimes B)^{\r a} \\
 &= \Bigl[ a\,b^2 + b\tr B_0\,X_1 + \bra\psi X_1 \ket\psi 
 + \tr Z_1\,(B_0\otimes B_0)^{\r s} +2\, \R\sqrt2\, b\, \bra\psi\varphi\rangle \\
 &\quad+ 2\, \R \bra{\psi\otimes\psi}\Phi\rangle 
 + 2\,\R \tr (\bra\psi \otimes B)^{\r s}\, Y^*_1 \Bigr]
 + \Bigl[ b\, \tr B_0\, X_2 - \bra\psi X_2 \ket\psi \\
 &\quad+ \tr (B_0\otimes B_0)^{\r a}\, Z_2 
 + 2\,\R \tr (\bra\phi \otimes B_0)^{\r a}\, Y_2^* \Bigr].
\end{align*}
We can restructure this expression
\begin{align*} 
 \tr A\, (B\otimes B)
 &= b\,a\,b + 2\,b\, \sqrt2\, \R\bra\varphi\psi\rangle + 2\,b\,
 \tr(\frac{X_1+X_2}{2})\, B_0 \\
 &\quad+ \bra\psi X_1 - X_2\ket\psi + 2\,\R \bra{\psi\otimes\psi}\Phi\rangle \\
 &\quad+ 2\,\R \tr (\bra\psi \otimes B)^{\r s}\, Y^*_1 
 + 2\,\R \tr(\bra\phi \otimes B_0)^{\r a}\, Y_2^* \\
 &\quad+ \tr (B_0\otimes B_0) \begin{pmatrix} Z_1&0 \\ 0&Z_2 \end{pmatrix}.
\end{align*}
We rewrite the first line of the right-hand side of the above equality. To make the 
link with $V_d(B)$ and $M_d(A)$, we express $\psi$ and $\varphi$ in their real and 
imaginary parts. We also use property~$(iii)$ of the map $V_d$. We then get
\begin{align*}
 b\,a\,b 
 &+ 2\,b\,\sqrt2\, \R\bra\varphi\psi\rangle + 2\,b \tr \Bigl( \frac{X_1+X_2}{2} \Bigr)\, 
 B_0 \\
 &= b\,a\,b + 2\,b\,\Bigl( \bra{\R\varphi}\sqrt2\, \R\psi\rangle 
 + \bra{\I\varphi\rangle}\sqrt2\, \I\psi\Bigr) + 2\,b\, 
 \Bigl\langle V_{d-1} \Bigl( \frac{X_1+X_2}{2} \Bigr) \,\Bigm|\,
 V_{d-1}(B_0) \Bigr\rangle.
\end{align*}
This looks promising, we can also try to express the second line in term of 
elements appearing in $V_d(B)$ and $M_d(A)$ or by looking at the real and imaginary 
part of the matrix- and vector components
\begin{align*}
 &\bra\psi X_1-X_2 \ket\psi + 2\,\R \bra{(\psi\otimes\psi)^{\r s}}\Phi\rangle \\
 &\quad= \sum_{i} \bigl( (\R\psi_i)^2 + (\I\psi_i)^2 \bigr) 
 \bigl( [X_1]_{ii} - [X_2]_{ii} \bigr) \\
 &\qquad+ 2\sum_{\{i,j \mid i<j\}} \Bigl[ \bigl(\R\psi_i\, \R\psi_j 
 + \I\psi_i\, \I\psi_j \bigr) \R \bigl( [X_1]_{ij} - [X_2]_{ij} \bigr) \\ 
 &\qquad\phantom{2\sum_{i<j} \Bigl[\ }- \bigl(\R\psi_i\, \I\psi_j + \I\psi_i\,\R\psi_j \bigr) 
 \I \bigl( [X_1]_{ij} - [X_2]_{ij} \bigr) \Bigr] \\
 &\qquad+ 2\sum_i \Bigl[ \bigl( (\R\psi_i)^2 - (\I\psi_i)^2 \bigr) \R\Phi_{ii}
 +2\, \R\psi_i\, \I\psi_i\, \I\Phi_{ii} \Bigr] \\
 &\qquad+ 2\sum_{\{i,j \mid i<j\}} \sqrt2 \Bigl[ \bigl(\R\psi_i\, \R\psi_j - 
 \I\psi_i\, \I\psi_j \bigr) 
 \R\Phi_{ij} + \bigl( \R\psi_i\, \I\psi_j + \I\psi_i\, \R\psi_j \bigr) \I\Phi_{ij} \Bigr] \\
 &\quad= \Bigl\langle \sqrt2\R\psi \,\Bigm|\, \frac{\R X_1 - \R X_2}{2} + 
 [\R\Phi] \,\Bigm|\, \sqrt2\R\psi \Bigr\rangle \\
 &\qquad+ \Bigl\langle \sqrt2 \I\psi \,\Bigm|\, \frac{\R X_1 - \R X_2}{2} - [\R\Phi]
 \,\Bigm|\, \sqrt2 \I\psi \Bigr\rangle \\
 &\qquad+ 2\Bigl\langle \sqrt2 \R\psi \,\Bigm|\, \frac{\I X_1 - \I X_2}{2} 
 + [\I\Phi] \,\Bigm|\, \sqrt2 \I\psi \Bigr\rangle.
\end{align*}
This also points out to the equality we are trying to prove. The fourth line is less 
straightforward but we can rewrite it
\begin{align*}
 &2\, \R \tr (\bra\psi \otimes B)^{\r s}\, Y^*_1 
 + 2\, \R \tr (\bra\phi \otimes B_0)^{\r a}\, Y_2^* \\
 &\quad= 2\R \sum_i \Bigl[ \sum_{k} \sqrt2\, [\bar{Y}_1]_{i,(k,k)} \psi_k\,[B_0]_{ik} \\
 &\qquad+ \sum_{\{k,\ell \mid k<\ell\}} \Bigl\{ \bigl( \psi_k\,[B_0]_{i\ell} 
 + \psi_\ell\, [B_0]_{ik} \bigr)
 [\bar{Y}_1]_{i,(k,\ell)} + \bigl( \psi_k\,[B_0]_{i\ell} - \psi_\ell\, [B_0]_{ik} \bigr) 
 [\bar{Y}_2]_{i,(k,\ell)} \Bigr\} \Bigr] \\
 &\quad= 2\, \R \sum_i \Bigl[ \sqrt2\, [\bar{Y}_1]_{i,(ii)} \psi_i\, [B_0]_{ii}
 + \sum_{\{k \mid k\neq i\}} \sqrt2\, [\bar{Y}_1]_{i,(k,k)} \psi_k\, [B_0]_{ik} \\
 &\qquad+ \sum_{\{k \mid k\neq i\}} \psi_k\, [B_0]_{ii} \bigl( [\bar{Y}_1]_{i,(i,k)} 
 + \epsilon_k^i [\bar{Y}_2]_{i,(i,k)} \bigr) \\
 &\qquad+ \sum_{\{k,\ell \mid k\neq \ell,\ \ell\neq i\}} \psi_k\, [B_0]_{i\ell} 
 \bigl( [\bar{Y}_1]_{i,(\ell,k)}
 + \epsilon_k^\ell [\bar{Y}_2]_{i,(\ell,k)} \bigr) \Bigr] \\
 &\quad= 2\, \R \sum_i \Bigl[ \sqrt2\, [\bar{Y}_1]_{i,(ii)} \psi_i
 +\sum_{\{k \mid k\neq i\}} \psi_k\, \bigl( [\bar{Y}_1]_{i,(i,k)} 
 + \epsilon_k^i[\bar{Y}_2]_{i,(i,k)} \bigr)\, [B_0]_{ii} \\
 &\qquad+ \sum_{\{\ell \mid \ell\neq i\}} \Bigl( \sqrt2\, [\bar{Y}_1]_{i,(\ell,\ell)} 
 \psi_\ell
 +\sum_{\{k \mid k\neq \ell\}} \psi_k\, \bigl( [\bar{Y}_1]_{i,(\ell,k)}
 + \epsilon_k^\ell [\bar{Y}_2]_{i,(\ell,k)} \bigr) \Bigr)\, [B_0]_{i\ell} \Bigr] \\
 &\quad= 2\,\R \sum_i \Bigl[ \sqrt2\, [\bar{Y}_1]_{i,(ii)} \psi_i
 + \sum_{\{k \mid k\neq i\}} \psi_k\, \bigl( [\bar{Y}_1]_{i,(i,k)}
 + \epsilon_k^i [\bar{Y}_2]_{i,(i,k)} \bigr) \, [B_0]_{ii} \\
 &\qquad+ \sum_{\{\ell \mid i<\ell\}} \Bigl( \sqrt2\, [\bar{Y}_1]_{i,(\ell,\ell)} 
 \psi_\ell
 + \sum_{\{k \mid k\neq \ell\}} \psi_k\, ([\bar{Y}_1]_{i,(\ell,k)} 
 + \epsilon_k^\ell [\bar{Y}_2]_{i,(\ell,k)} \bigr) \Bigr)\,[B_0]_{i\ell} \\
 &\qquad+ \sum_{\{\ell \mid i<\ell\}} \Bigl( \sqrt2\, [\bar{Y}_1]_{\ell,(i,i)} \psi_i
 + \sum_{\{k \mid k\neq i\}} \psi_k\, ([\bar{Y}_1]_{\ell,(i,k)}
 + \epsilon_k^i  [\bar{Y}_2]_{\ell,(i,k)} \bigr) \Bigr)\, [\bar{B_0}]_{i\ell} \Bigr] \\
 &\quad= 2\, \sum_i \Bigl[ \bigl( \sqrt2\, \R[Y_1]_{i,(ii)}\, \R\psi_i
 + \sqrt2\, \I[Y_1]_{i,(ii)}\, \I\psi_i \bigr) \\
 &\qquad+ \sum_{\{k \mid k\neq i\}} \Bigl( \R\psi_k\, \R \bigl( [Y_1]_{i,(i,k)}
 + \epsilon_k^i\, [Y_2]_{i,(i,k)} \bigr)
 + \I\psi_k\, \I \bigl( [Y_1]_{i,(i,k)} \\
 &\qquad\phantom{\sum_{\{k \mid k\neq i\}} \Bigl(}+ 
 \epsilon_k^i\, [Y_2]_{i,(i,k)} \bigr) \Bigr)\, [B_0]_{ii} \\
 &\qquad+ \sum_{\{\ell \mid i<\ell\}} \Bigl\{ \Bigl( \sqrt2\, \R[Y_1]_{i,(\ell,\ell)} 
 \R\psi_\ell
 + \sqrt2\, \I[Y_1]_{i,(\ell,\ell)} \I\psi_\ell
 + \sqrt2\, \R[Y_1]_{\ell,(i,i)} \R\psi_i \\
 &\qquad\phantom{\sum_{\{\ell \mid i<\ell\}} \Bigl(}+ \sqrt2\, 
 \I[Y_1]_{\ell,(i,i)} \I\psi_i \Bigr) \\
 &\qquad+ \sum_{\{k \mid k\neq \ell\}} \Bigl( \R\psi_k\, \R \bigl([Y_1]_{i,(\ell,k)}
 + \epsilon_k^\ell\, [Y_2]_{i,(\ell,k)} \bigr)
 + \I\psi_k\, \I \bigl( [Y_1]_{i,(\ell,k)}
 + \epsilon_k^\ell\, [Y_2]_{i,(\ell,k)} \bigr) \Bigr) \\
 &\qquad+ \sum_{\{k \mid k\neq i\}} \Bigl( \R\psi_k\, \R \bigl( [Y_1]_{\ell,(i,k)}
 + \epsilon_k^i\, [Y_2]_{\ell,(i,k)} \bigr)
 + \I\psi_i\, \I \bigl( [Y_1]_{\ell,(i,k)} \\
 &\qquad\phantom{\sum_{\{k \mid k\neq i\}} \Bigl(}+ 
 \epsilon_k^i\, [Y_2]_{\ell,(i,k)} \bigr) \Bigr) \R [B_0]_{i\ell} \Bigr\} \\
 &\qquad+ \sum_{\{\ell \mid i<\ell\}} \Bigl\{ \Bigl( -\sqrt2\, 
 \R[Y_1]_{i,(\ell,\ell)} \I\psi_l
 + \sqrt2\, \I[Y_1]_{i,(\ell,\ell)} \R\psi_\ell
 + \sqrt2\, \R[Y_1]_{\ell,(i,i)} \I\psi_i \\
 &\qquad\phantom{\sum_{\{\ell \mid i<\ell\}} \Bigl(}- \sqrt2\, 
 \I[Y_1]_{\ell,(i,i)} \R\psi_i \Bigr) \\
 &\qquad+ \sum_{\{k \mid k\neq \ell\}} \Bigl( -\I\psi_k\, \R \bigl( [Y_1]_{i,(\ell,k)}
 + \epsilon_k^\ell\, [Y_2]_{i,(\ell,k)} \bigr)
 + \R\psi_k\, \I \bigl( [Y_1]_{i,(\ell,k)}
 + \epsilon_k^\ell\, [Y_2]_{i,(\ell,k)} \bigr) \Bigr) \\
 &\qquad+ \sum_{\{k \mid k\neq i\}} \Bigl( \I\psi_k\, \R \bigl( [Y_1]_{\ell,(i,k)}
 + \epsilon_k^i\, [Y_2]_{\ell,(i,k)} \bigr)
 - \R\psi_k\, \I \bigl( [Y_1]_{\ell,(i,k)} \\
 &\qquad\phantom{\sum_{\{k \mid k\neq i\}} \Bigl(}+ 
 \epsilon_k^i\, [Y_2]_{\ell,(i,k)} \bigr) \Bigr) \I [B_0]_{i\ell} \Bigr\} \Bigr] \\
 &\quad= 2\,\Bigl\langle \sqrt2 \R\psi \,\Bigm|\, T_1 (Y_1 , Y_2) \,\Bigm|\,
 V_{d-1}(B_0) \Bigr\rangle + 2\,\Bigl\langle \sqrt2 \I\psi \,\Bigm|\, T_2 (Y_1 , Y_2)
 \,\Bigm|\, V_{d-1}(B_0) \Bigr\rangle.
\end{align*}
Finally, using induction on $d$
\begin{equation*}
 \tr (B_0\otimes B_0) \begin{pmatrix} Z_1&0 \\ 0&Z_2 \end{pmatrix}
 = \Bigl\langle V_{d-1}(B_0) \,\Bigm|\, 
 M_{d-1} \Bigl( \begin{pmatrix} Z_1&0 \\ 0&Z_2 \end{pmatrix} \Bigr) \,\Bigm|\, V_{d-1}(B_0)
 \Bigr\rangle.
\end{equation*}
If we put this all together we get what we wanted to prove, namely
\begin{align*}
 &\tr A\,(B\otimes B) \\
 &\quad= b\,a\,b + 2\,b\, \bigl( \bra{\R\varphi} \sqrt2\, \R\psi\rangle
 + \bra{\I\varphi} \sqrt2\, \I\psi \rangle \bigr) \\ 
 &\qquad+ 2\,b\, \Bigl\langle V_{d-1} \Bigl( \frac{X_1+X_2}{2} \Bigr) \,\Bigm|\, 
 V_{d-1}(B_0) \Bigr\rangle \\
 &\qquad+ \Bigl\langle \sqrt2 \R \psi \,\Bigm|\, \frac{\R X_1 - \R X_2}{2} + [\R\Phi] 
 \,\Bigm|\, \sqrt{2}\R \psi \Bigr\rangle \\
 &\qquad+ \Bigl\langle \sqrt2 \I\psi \,\Bigm|\, \frac{\R X_1 - \R X_2}{2} -[\R\Phi] 
 \,\Bigm|\, \sqrt2 \I\psi \Bigr\rangle \\
 &\qquad+ 2 \Bigl\langle \sqrt2 \R\psi \,\Bigm|\, \frac{\I X_1 - \I X_2}{2} 
 + [\I\Phi] \,\Bigm|\, \sqrt2 \I\psi \Bigr\rangle \\
 &\qquad+ 2\,\bra{\sqrt2 \R\psi} T_1 (Y_1 + Y_2) \ket{V_{d-1}(B_0)}
 +2\, \bra{\sqrt2 \I\psi} T_2 (Y_1 + Y_2) \ket{V_{d-1}(B_0)} \\
 &\qquad+ \bra{V_{d-1}(B_0)} M_d \Bigl( 
 \begin{pmatrix} Z_1&0 \\ 0&Z_2 \end{pmatrix} \Bigr) \ket{V_{d-1}(B_0)} \\
 &\quad= \bra{V_d(B)} M_d(A) \ket{V_d(B)}.
\end{align*}
To summarize, we have found maps 
\begin{equation*}
 V_d: \c M^{\r h}_d(\Cx) \rightarrow \Rl^{d^2}
 \qquad\text{and}\qquad 
 M_d: \Bigl( \c M_d(\Cx) \otimes \c M_d(\Cx) \Bigr)^{\r h} \rightarrow 
 \c M^{\r h}_{d^2}(\Rl)
\end{equation*}
with properties that allow us to prove the second part of theorem~\ref{characterisation B-E}, 
see section~\ref{The quantum case}.
\bigskip

\noindent
\textbf{Acknowledgements:}
This work was partially supported by F.W.O., Vlaanderen grant G.0109.01.


\begin{thebibliography}{99}
\bibitem{Fin}
B.~de~Finetti, 
La pr\'evision: ses lois logiques, ses sources subjectives, 
\textit{Ann.\ Inst.\ H.~Poincarr\'e} \textbf{7}, 1--68 (1937)
\bibitem{Duk}
J.~Dukelsky, R.~Mosseri and J.~Vidal, 
Entanglement in a first order phase transition, 
\textit{Phys.\ Rev.\ A} \textbf{69}, 054101 (2004)
and
R.~Mosseri, G.~Palacios and J.~Vidal, 
Entanglement in a second-order quantum phase transition, 
\textit{Phys.\ Rev.\ A} \textbf{69}, 022107 (2004)
\bibitem{Fuc}
C.A.~Fuchs, R.~Schack and P.F.~Scudo,
A de~Finetti representation theorem for quantum process tomography,
\textit{Phys.\ Rev.\ A} \textbf{69} 062305/1--6 (2004)
\bibitem{Fuk}
M.~Fukuda and A.S.~Holevo, 
On Weyl-covariant channels, 
quant-ph/0510148 (2006) 
\bibitem{Ham}
M.~Hamermesh,
\textit{Group Theory and its Applications to Physical Problems},
Addison-Wesley, Reading MA. (1962)
\bibitem{Hew}
E.~Hewitt and L.J.~Savage, 
Symmetric measures on Cartesian products, 
\textit{Trans.\ Amer.\ Math.\ Soc.\ }\textbf{80}, 470--501 (1955)
\bibitem{Hud}
R.L.~Hudson and G.R.~Moody, 
Locally normal symmetric states and a analogue of de Finetti's theorem, 
\textit{Z.\ Wahrschein.\ verw.\ Geb. }\textbf{33} 343--351 (1976)
\bibitem{Kin}
C.~King and M.B.~Ruskai, 
Minimal entropy of states emerging from noisy quantum channels, 
\textit{IEEE Trans.\ Inf.\ Theory (USA)}\textbf{47}, 192--209 (2001)
\bibitem{Koe}
R.~K\"onig and R.~Renner, 
A de~Finetti representation for finite symmetric quantum states, 
\textit{J.\ Math.\ Phys.\ }\textbf{46}, 122108 (2005) 
\bibitem{Nie}
M.A.~Nielsen and I.L.~Chuang,  
\textit{Quantum Computation and Quantum Information},
Cambridge University Press, Cambridge (2000)
\bibitem{Rei}
L.E.~Reichl,  
\textit{A Modern Course in Statistical Physics},  
John Wiley \& Sons, New York, 2nd edition (1998)
\bibitem{Sto}
E.~St\o rmer, 
Symmetric states of infinite tensor product of $C^*$-algebras, 
\textit{J.\ Funct.\ Anal.\ }\textbf{3}, 48 (1969)
\bibitem{Wer}
R.F.~Werner,
Remarks on a quantum state extension problem,
\textit{Lett.\ Math.\ Phys.\ }\textbf{19}, 319--326 (1990)
\end{thebibliography}
\end{document}